\pdfoutput=1
\documentclass{article}
\usepackage{spconf,amsmath,graphicx}
\usepackage{amsfonts}
\usepackage{amssymb}
\usepackage{algorithm}
\usepackage[noend]{algpseudocode}
\usepackage{multirow}
\usepackage{multicol}
\usepackage{tabularx}
\usepackage{enumitem}
\usepackage[dvipsnames, table]{xcolor}
\usepackage{float}
\usepackage{url}
\usepackage{hyperref}
\usepackage{soul}
\usepackage{changepage,threeparttable} 
\usepackage{arydshln}
\usepackage{cleveref}
\usepackage{caption}

\usepackage{booktabs}
\usepackage[disable]{todonotes}

\makeatletter
\newcommand*\iftodonotes{\if@todonotes@disabled\expandafter\@secondoftwo\else\expandafter\@firstoftwo\fi}  
\makeatother

\newcommand{\note}[4][]{\todo[author=#2,color=#3,size=\scriptsize,fancyline,inline,caption={},#1]{#4}} 
\newcommand*{\genbf}[1]{\ifmmode\mathbf{#1}\else\textbf{#1}\fi}

\newcommand{\wer}{\text{WER}}
\newcommand{\ad}[1][A]{\mathcal{#1}}
\DeclareMathOperator{\ctc}{CTC}
\DeclareMathOperator{\kldiv}{KL-div}
\DeclareMathOperator{\softmax}{softmax}

\newcommand{\anuj}[2][]{\note[#1]{anuj}{cyan!40}{#2}}

\newcommand{\benchmark}{LibriContinual}
\newcommand{\benchmarkshort}{LC}
\newcolumntype{H}{>{\lrbox0}c<{\endlrbox}@{}}

\makeatletter
\def\adl@drawiv#1#2#3{%
        \hskip.5\tabcolsep
        \xleaders#3{#2.5\@tempdimb #1{1}#2.5\@tempdimb}%
                #2\z@ plus1fil minus1fil\relax
        \hskip.5\tabcolsep}
\newcommand{\cdashlinelr}[1]{%
  \noalign{\vskip\aboverulesep
           \global\let\@dashdrawstore\adl@draw
           \global\let\adl@draw\adl@drawiv}
  \cdashline{#1}
  \noalign{\global\let\adl@draw\@dashdrawstore
           \vskip\belowrulesep}}
\makeatother

\title{Continual Learning for On-Device Speech Recognition \\ using Disentangled Conformers}
\name{\begin{tabular}{@{}c@{}}
Anuj Diwan\thanks{*Work done at Meta Inc.}$^{*,1}$, Ching-Feng Yeh$^{2}$, Wei-Ning Hsu$^{2}$, Paden Tomasello$^{2}$,\\ 
Eunsol Choi$^{1}$, David Harwath$^{1}$, Abdelrahman Mohamed$^{2}$
\end{tabular}}

\address{$^1$ University of Texas at Austin \qquad $^2$ Meta Inc. \\ \texttt{\{anuj.diwan,eunsol,harwath\}@utexas.edu} \\ \texttt{\{cfyeh,wnhsu,padentomasello,abdo\}@meta.com}}

\begin{document}
%
\maketitle
\begin{abstract}
Automatic speech recognition research focuses on training and evaluating on static datasets. Yet, as speech models are increasingly deployed on personal devices, such models encounter user-specific distributional shifts.
To simulate this real-world scenario, we introduce \benchmark{}, a continual learning benchmark for speaker-specific domain adaptation derived from LibriVox audiobooks, with data corresponding to $118$ individual speakers and $6$ train splits per speaker of different sizes. Additionally, current speech recognition models and continual learning algorithms are not optimized to be compute-efficient. We adapt a general-purpose training algorithm NetAug for ASR and create a novel Conformer variant called the DisConformer (Disentangled Conformer). This algorithm produces ASR models consisting of a frozen `core' network for general-purpose use and several tunable `augment' networks for speaker-specific tuning. Using such models, we propose a novel compute-efficient continual learning algorithm called DisentangledCL. Our experiments show that the DisConformer models significantly outperform baselines on general ASR i.e. LibriSpeech ($15.58\%$ rel.\@ WER on test-other). On speaker-specific LibriContinual they significantly outperform trainable-parameter-matched baselines (by $20.65\%$ rel.\@ WER on test) and even match fully finetuned baselines in some settings.

\end{abstract}
\begin{figure}[t]
    \centering
    \includegraphics[width=\columnwidth]{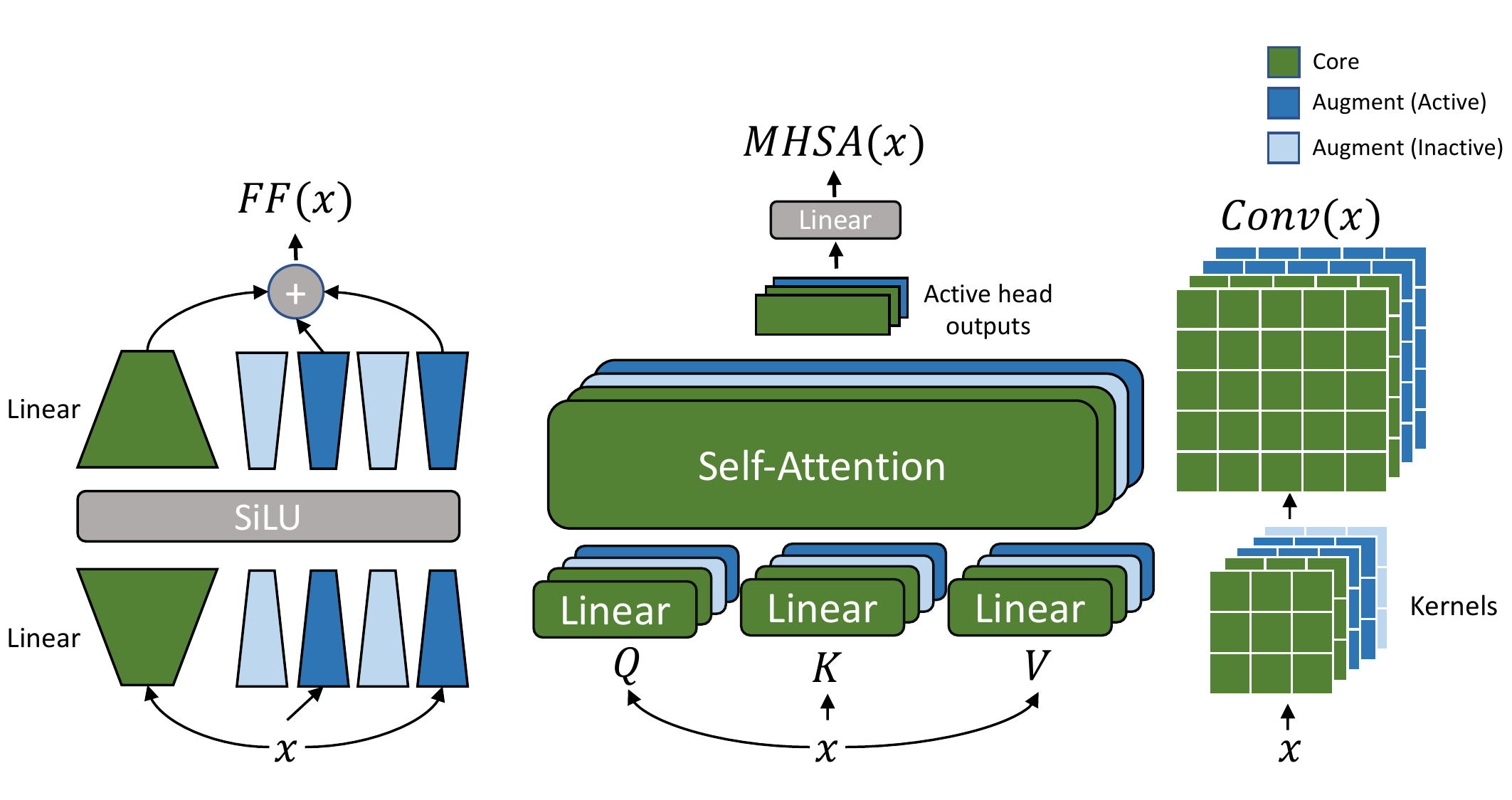}
    \caption{The DisConformer architecture depicting disentanglement in the Feedforward, Self-Attention and Convolution modules.}
    \label{fig:mainfig}
\end{figure}
\section{Introduction}
\label{sec:intro}
Today, speech recognition models are deployed on millions of personal devices.
Such deployed models encounter an ever-changing distributional shift associated with their user's environment (e.g. speaker characteristics).
Models should continually learn and adapt to their environment in a tractable, compute-efficient manner.
While doing so, models should still perform well for other speakers without suffering from catastrophic forgetting~\cite{MCCLOSKEY1989109}.
Measuring such a continual-learning ability is not possible with current static ASR datasets. Therefore, we introduce the \textbf{\benchmark{}} benchmark, a continual learning dataset for speaker-specific adaptation. This new benchmark is derived from LibriVox audiobooks and consists of training, validation and test datasets corresponding to $118$ different speakers, with $6$ training splits per speaker ranging from $10$ min to $10$ hr of speaker-specific data. Our benchmark measures the ability of models to continually adapt to new speakers in a compute-efficient manner, while preserving performance on the training dataset.
We describe the \benchmark{} benchmark in Section~\ref{sec:benchmark}.

Furthermore, current speech recognition models do not inherently support compute-efficient techniques for on-device continual learning. Current continual learning techniques for ASR from prior work~\cite{https://doi.org/10.48550/arxiv.2104.01616, https://doi.org/10.48550/arxiv.2112.09427, Sadhu2020} require finetuning the entire model, which is not compute efficient.
We propose (a) a novel general-purpose ASR algorithm derived from NetAug~\cite{cai2021netaug} to train ASR models that consist of `core' and `augment' networks and (b) \textbf{DisentangledCL}, a novel continual learning algorithm inspired by adapter networks~\cite{https://doi.org/10.48550/arxiv.1902.00751} that only requires finetuning a small subset of these `augment' networks and is compute-efficient. We apply our disentanglement approach to the Conformer~\cite{gulaticonformer2020} to obtain \textbf{DisConformers}. We describe the DisConformer and DisentangledCL in Section~\ref{sec:approach}.

We find DisConformer models significantly outperform baselines on speaker-independent LibriSpeech by $15.58\%$ relative WER on test-other with n-gram LM decoding; further, on speaker-specific LibriContinual, they significantly outperform trainable-parameter-matched baselines (by $20.65\%$ relative on test set with n-gram LM) and sometimes even match fully finetuned baselines (in the DisConformer-Att and -Conv settings), while finetuning ${<}13\%$ of parameters.

\section{Related Work}
\noindent \textbf{Continual Learning for Speech.} \cite{https://doi.org/10.48550/arxiv.2104.01616, https://doi.org/10.48550/arxiv.2112.09427, Sadhu2020} all explore continual learning in the context of ASR using regularization-based (e.g. EWC~\cite{Kirkpatrick_2017})) and data-replay based (e.g. GEM~\cite{https://doi.org/10.48550/arxiv.1706.08840}) approaches. Other work explores settings such as SSL~\cite{https://doi.org/10.48550/arxiv.2107.13530} and online learning~\cite{https://doi.org/10.48550/arxiv.2207.05071}.

\noindent \textbf{Disentangled Models and On-Device ASR.} Our DisConformer is trained using the NetAug algorithm~\cite{https://doi.org/10.48550/arxiv.2110.08890} proposed for CNNs which we adapt for Transformers and ASR. While the original NetAug paper only uses the `core' network at inference and discards the `augment' networks,  we repurpose and use the `augment' networks for performing disentangled continual learning. For on-device ASR, other prior work such as~\cite{https://doi.org/10.48550/arxiv.2203.15610, https://doi.org/10.48550/arxiv.2110.08352} train several subnets within a network to decrease model size while preserving high accuracies.

\section{\benchmark{}: A Continual Learning Benchmark}
\label{sec:benchmark}
Real-world speech models encounter user-specific distributional shift and must adapt to this domain shift. To measure this ability, we present the \benchmark{} benchmark, a continual learning speaker adaptation benchmark. The same model should be capable of efficient speaker adaptation while still maintaining general-purpose ASR performance (e.g. to transcribe audio not spoken by the user like videos, phone calls, etc.). Our evaluation framework reflects these three requirements: a) efficient adaptation b) high speaker performance and c) high general-purpose performance.

\subsection{Dataset Creation}
\benchmark{} is sourced from the LibriVox project: open-sourced speech from thousands of open-domain audiobooks. 
We first remove speakers already in the Librispeech~\cite{panayotov2015librispeech} dataset. Then, we select a subset of the remaining speakers that have at least $2$-hrs of data and $2$ audiobooks each to make val and test sets and at least $10$-hr of data to create a training set. Thus, we select a subset of $118$ speakers that have sufficient data in order to create a $10$-hr training set and validation and test sets of at least $2$-hr, ensuring that there is no overlap between the audiobooks used in each set. We apply a Voice Activity Detector (VAD) to segment each audiobook into utterances of max duration $16$ s. Finally, subsets of the $10$ hr training set are constructed to obtain $5$ hr, $2$ hr, $1$ hr, $30$ min and $10$ min training splits, such that each split is a superset of the next split.
We obtain synthetic text transcriptions by running ASR using a wav2vec 2.0 Large~\cite{baevski2020wav2vec} model
pretrained and self-trained on LibriLight~\cite{librilight} and finetuned on Librispeech~\cite{panayotov2015librispeech} and decode with a word Transformer LM.
\footnote{beam=100;beamthres=20;lmweight=1.51;wordscore=2.06;silweight=-3} 
Since the transcriptions are not human-derived, progress on this benchmark 
should only be interpreted as making better wav2vec2.0 model-like predictions.

Table~\ref{tab:datasetstats} contains the \benchmark{} dataset statistics, with information about the number of hours and utterances per speaker for each split. While train set durations are fixed (e.\@g.\@ 10h) and have nearly no variance across speakers, the val and test sets of each speaker have variable durations (2-14h).

\begin{table}[!htp]\centering
\begin{tabular}{cccc}\toprule
Subset &\#hrs/spkr &\#utts/spkr \\\midrule
train-10min &0.17 ± 0.001 &114 ± 28 \\
train-30min &0.50 ± 0.001 &337 ± 81 \\
train-1hr &1.00 ± 0.001 &677 ± 163 \\
train-2hr & 2.00 ± 0.001 &1356 ± 322 \\
train-5hr &5.00 ± 0.003 &3387 ± 806 \\
train-10hr &10.00 ± 0.005 &6772 ± 1608 \\
valid &3.13 ± 1.86 &2125 ± 1406 \\
test &2.66 ± 1.15 &1880 ± 1101 \\
\bottomrule
\end{tabular}
\caption{\benchmark{} dataset statistics. For both \# hrs/spkr and \# utts/spkr, mean and standard deviation across speakers is reported.}\label{tab:datasetstats}
\end{table}

\subsection{Evaluation Framework}
\label{subsec:eval}
Given a general ASR model $\ad[M]$ trained on an ASR dataset $\ad[D]_{orig}$ (LibriSpeech in all experiments) and a continual learning algorithm $\ad(\ad[M],\ad[D])$ to finetune it on a dataset $\ad[D]$, we run $\ad$ on $\ad[M]$ for every speaker $s$ to obtain $118$ speaker-specific models $\ad[M]^{(s)} = \ad(\ad[M],\ad[D]^{(s)}_{\benchmarkshort{}, train})$, where $\ad[D]^{(s)}_{\benchmarkshort{}, train}$ is the \benchmark{} (\benchmarkshort{}) train data for speaker $s$. We report:
    
\noindent \textbf{\textit{Number of trainable params}} \underline{\#CL-Params} available during continual learning, a proxy for measuring compute-efficiency of $\ad$.

\noindent \textbf{\textit{Speaker-aggregate}} $\underline{\wer_{\benchmarkshort{}}}$. Each model $\ad[M]^{(s)}$ is evaluated on $s$'s val/test sets $\ad[D]^{(s)}_{\benchmarkshort{}, val/test}$ 
    to compute $118$ different WERs $\wer^{(s)}_{\benchmarkshort{}}$. Their median is taken to define a single number, $\wer_{\benchmarkshort{}}$. 
    
    
\noindent \textbf{\textit{Original-aggregate}} $\underline{\wer_{orig}}$. Each
    model $\ad[M]^{(s)}$ is evaluated on the $\ad[D]_{orig}$ test set to obtain $118$ different WERs and then their median is taken to compute $\wer_{orig}$, measuring the ability to retain performance on the original $\ad[D]_{orig}$.

The above evaluation is repeated for every train split.
Our benchmark contains data for $6$ train splits but in our experiments we only report results for the $1$ hr and $10$ hr splits to be concise.


\section{The DisConformer Model}
\label{sec:approach}
We propose the DisConformer model (Fig~\ref{fig:mainfig}) based on a \textit{disentangled} approach designed to achieve a good tradeoff between adapting to new speakers and minimizing catastrophic forgetting by training two different types of model parameters: `core' $W_c$ and `augment' $W_a$. Given an input $x$, the parameters used for the forward pass are dynamically constructed from $W_c$ and a (potentially random) subset of $W_a$. Given a (randomized) `selector' function $S(W_a, x) \subseteq W_a$, the forward pass uses $W_c$ and $S(W_a, x)$: $\ad[M]([W_c, S(W_a, x)], x)$. The core is always active while only a subset of augment params are. Then, the core is used for general-purpose ASR while the augment params are finetuned on speaker-specific data.

Our approach can be applied to any neural network, but we focus on the Conformer~\cite{gulaticonformer2020} model. We dub these versions as `DisConformers' and propose disentangling the three types of modules (Feedforward, Self-Attention and Convolution), giving rise to DisConformer-FF, -MHSA, and -Conv. For e.g., in DisConformer-FF, the FF module is disentangled while the MHSA and Conv modules only have core parameters like a standard Conformer.



\textbf{DisConformer-FF}: The FF module in a Conformer consists of a sequence of layers: a linear layer, a non-linearity, and another linear layer.
    In the DisConformer-FF, we disentangle the feedforward dimension $f$ into core and augment dimensions. The first linear layer has a core module with parameters $W_{1, c} \in \mathbb{R}^{d \times f_c}, b_{1, c} \in \mathbb{R}^{d \times f_c}$ and $n_a$ augment \textit{experts}, each with parameters $W^{i}_{1, a} \in \mathbb{R}^{d \times f_a}, b^{i}_{1, a} \in \mathbb{R}^{d \times f_a}$, where $f= f_c + n_af_a$ is the feedforward dimension in the vanilla Conformer. Similarly, the second linear layer has a core module with weight parameters $W_{2, c} \in \mathbb{R}^{f_c \times d}$ and $n_a$ augment experts with parameters $W^{i}_{2, a} \in \mathbb{R}^{f_a \times d}, b_{2} \in \mathbb{R}^d$. Given an input $x$ and a subset of $r$ active augment experts with indices $i_1, i_2, \ldots , i_r$, the output $y$ is computed as in \crefrange{eq:ff1}{eq:ff2}:
    \begin{align}
    \label{eq:ff1}
        h_c &= \sigma(W_{1,c}x + b_{1,c}) \\
        h_a &= \sigma([W^{i_1}_{1,a},\ldots,W^{i_r}_{1,a}]x + [b^{i_1}_{1,a},\ldots,b^{i_r}_{1,a}]) \\
        y &= W_{2,c}h_c +[W^{i_1}_{2,a},\ldots,W^{i_r}_{2,a}]h_a + b_2 \label{eq:ff2}
    \end{align}
    
\textbf{DisConformer-Att}: The Att module in a Conformer performs multi-head self-attention with $h$ different heads. In DisConformer-Att, we first disentangle the heads into $h_c$ core heads and $h_a$ augment heads, where $h = h_c+h_a$. Given a subset of $r$ active augment heads, we perform multi-head self-attention as usual, but using just the $h_c$ core heads and $r$ augment heads, not all the $h_a$ augment heads. Formally, each head $i$ has self-attention projection weights $W^Q_i \in \mathbb{R}^{d \times d_q}, W^K_i \in \mathbb{R}^{d \times d_k}, W^V_i \in \mathbb{R}^{d \times d_v}$ and output projection weights $W^O_i \in \mathbb{R}^{d_v \times d}$ for query, key, and value dimensions $d_q,d_k,d_v$. Given an input $x$ and a subset of $r$ active augment experts with indices $S_a = \{i_1, i_2, \ldots , i_r\}$, the output $y$ is computed as in \crefrange{eq:att1}{eq:att2}:
\begin{align}
    \label{eq:att1}
        Q &= K = V = x \\
        y_i &= \text{Attention}(QW^Q_i, KW^K_i, VW^V_i) \forall i \in \{1,\ldots,h\} \\
        y &= \sum_{i=1}^{h_c}{y_i W^O_i} + \sum_{i \in S_a}{y_i W^O_i}
\label{eq:att2}
\end{align}

\textbf{DisConformer-Conv}: The Conv module in a standard Conformer consists of a sequence of layers: a Pointwise Conv $PC_1$, a 1D Depthwise Conv $DC$, Layer Norm $LN$, another Pointwise Conv $PC_2$. Each layer is parametrized by the number of intermediate conv channels, $d_{conv}$. For e.\@g.\@, $PC_1$ maps the input from $d$ to $d_{conv}$ channels.
    In the DisConformer-Conv, we disentangle the $d_{conv}$ channels into $d_c$ core channels and $d_a$ augment channels. Given a subset of $r$ active augment channels, we index into each layer's kernels to create new kernels with $d_{conv}' = d_c+r$ intermediate channels and compute convolutional operations normally using this new kernel.



\subsection{General ASR Training using NetAug}
\label{subsec:initialasrdescription}
We train the DisConformer as a general-purpose ASR model on $\ad[D]_{orig}$ i.e. Librispeech, using \textit{NetAug training} inspired by~\cite{cai2021netaug}.
Let the DisConformer-FF (/Att/Conv) model have $n_{ffn}$ (/$n_{att}$/$n_{conv}$) augment experts (/heads/channels). For ease of explanation, we describe the approach using DisConformer -FF but the approach is analogously applied to DisConformer-Att and DisConformer-Conv. Given a training example $(x,y)$, we first uniformly sample a number $n$ from $\{1,2,4,\ldots,n_{ffn}\}$. Then, we uniformly sample $n$ FF augment experts from the total $n_{ffn}$ experts, whose parameters one can denote as $W_{aug, ffn}$. That is, we sample a random-sized random subset of augment params. Denoting the core parameters by $W_{core}$, we can define the training loss $L(\ad[M], x, y)$ as in~\cref{eq:netaug}:
\begin{align}
\label{eq:netaug}
&~ L(\ad[M], x, y) = \ctc(\ad[M](W_{core}, x),y) \\ +
&~ \alpha \ctc(\ad[M]([W_{core}, W_{aug, ffn}], x), y) \nonumber
\end{align}
where $\ctc$ is the Connectionist Temporal Classification loss~\cite{graves2006connectionist}) and $\alpha$ is a hyperparameter; in practice, we always set it to $1.0$ as that performed best on the Librispeech dev-other validation set in initial experiments. This loss encourages the model to train the core parameters in isolation (term 1) as well as in conjunction with a random subset of augment parameters (term 2).

\anuj{Is figure required here?}

\subsection{Continual Learning using DisentangledCL}
\label{subsec:cl}
We introduce a novel compute-efficient continual learning algorithm \textit{DisentangledCL}. Again, for brevity, we describe the approach using DisConformer-FF. We first start with a general-purpose ASR model trained using NetAug.
To finetune on a training dataset $\ad[D]$, we randomly select a subset of $k_{ffn} {<}\ n_{ffn}$ augment experts, denoting their params by $W^{k}_{aug,ffn}$, such that $|W^{k}_{aug,ffn}| << |W_{core}|$ i.e no. of trainable augment params is a small fraction of core params; at most $13\%$ in all experiments. 
We then finetune these $W^{k}_{aug,ffn}$ parameters while $W_{core}$ is frozen, using the regular CTC loss $\ctc(\ad[M]([W_{core}, W^{k}_{aug,ffn}], x), y)$. 
We use $W_{core}$ i.e. the core parameters for general-purpose inference (on $\ad[D]_{orig}$). For speaker-specific inference, we use $[W_{core}, W^{k}_{aug,ffn}]$. Thus, we get the best of both worlds; performance is retained on the original dataset via the core parameters, while speaker-specific improvements can come from the finetuned augment parameters.
\section{Experiments}
\subsection{Experimental Setup}
\label{sec:expsetup}
We use the standard Conformer~\cite{gulaticonformer2020} architecture, but with a time reduction layer (similar to~\cite{https://doi.org/10.48550/arxiv.1811.06621, chanlisten2016}) instead of $2$ learnable CNN layers for efficiency.
All models share the following hyperparams: $256$ model dim, $30$ output dim, $16$ layers, $64$ FF dim per expert, $31$ depthwise conv kernel, and $0.1$ dropout. The aspects in which they differ are summarized in Table~\ref{tab:archs}.
The output vocabulary consists of the English alphabet ($26$ letters), space, apostrophe and CTC blank.
\begin{table}[h!]
    \centering
    \scriptsize
    \resizebox{\linewidth}{!}{%
    \begin{tabular}{cc c c}
    \toprule
        & DisCo-FF & DisCo-Att & DisCo-Conv \\
        \midrule
        \# FF (core,aug) & (8,12) & (20,0) & (20,0) \\
        \# Att (core,aug) heads & (4,0) & (2,2) & (4,0) \\
        Conv channels/expert & 16 & 16 & 8 \\
        \# Conv (core,aug) & (16,0) & (16,0) & (16,16) \\
    \bottomrule
    \end{tabular}}
    \caption{Summary of DisConformer model architectures.}
    \label{tab:archs}
\end{table}

\textbf{NetAug ASR Training: Details.}
We train on the Librispeech \cite{panayotov2015librispeech} 960-hr training set. We use SpecAugment with $2$ $27$-channel freq masks and $2$ $100$-frame time masks. 
 We use Adam with an lr of $0.0004$,$\beta_1{=}0.9, \beta_2{=}0.98$ and train for $200$k steps on $16$ GPUs with a $4$-stage linear LR schedule: warmup $8\%$, const $32\%$, decay $40\%$, const $20\%$. We use a per-GPU batch size of $32$ subject to a max of $320$ s. We choose the checkpoint with the min WER on Librispeech dev-clean + dev-other.

\textbf{DisentangledCL: Details.}
We set $k_{ffn} {=} 2$ for DisCo-FF, $k_{att} {=} 2$ for DisCo-Att, and $k_{conv} {=} 12$ for DisCo-Conv. We use Adam with an lr of $0.0001$, $\beta_1{=}0.9, \beta_2{=}0.98$. In this paper, we run experiments for only the $1$ hr and $10$ hr subsets. For $10$ hr, we train for $30$k steps while for the $1$ hr subset, we train for $10$k steps on $1$ GPU. These numbers were chosen to ensure overall model convergence. We use the same $3$-stage LR schedule for both; $40\%$ const, $40\%$ decay, $20\%$ const. Other hyperparams are same as NetAug training. We report eval results using both Viterbi decoding and n-gram LM decoding. We use a $4$-gram LM trained on the Librispeech book corpus with beam=$20$, lmweight=$1.74$, wordscore=$-0.52$.
\subsection{Baselines}
\hspace{\parindent}\textbf{Baseline Models:} For each of the three DisConformer models, we construct corresponding Conformer baselines dubbed Base-FF, Base-Att and Base-Conv. Base-FF is a Conformer with an FFN dimension of $512 = 64 \times 8$ and otherwise identical to DisCo-FF. Thus, it has the same architecture as a DisCo-FF with only its core ($8$ experts each with dim $64$). Similarly, Base-Att is a Conformer with $2$ heads (identical to DisCo-Att with only its $2$ core heads) and Base-Conv is a Conformer with $128 = 8 \times 16$ channels (identical to DisCo-Conv with only its $16$ core experts).
We perform general-purpose ASR training on Librispeech using the regular $\ctc$ loss with the same optimizer hyperparams and number of steps as the DisConformers and choose the best-performing model on Librispeech dev-clean + dev-other.

\textbf{Baseline Continual Learning algorithms:} All ASR continual learning techniques investigated in prior work~\cite{https://doi.org/10.48550/arxiv.2104.01616, https://doi.org/10.48550/arxiv.2112.09427, Sadhu2020} finetune the entire model, which is more computationally expensive than our DisentangledCL which only finetunes a small subset of parameters. We first investigate two existing baseline CL algorithms (which finetune the whole model). Further, for a fairer comparison with our approach, we analyze simple, efficient variants of both algorithms.

\noindent \underline{(1) Full-FT}: We fully finetune the baseline models using CTC loss. We use the same hyperparameters as the DisConformers, except that we use a more stable learning rate of $0.00005$ for the $1$ hr subset.

\noindent \underline{(2) KD (Knowledge Distillation)}:
Following previous work~\cite{https://doi.org/10.48550/arxiv.2104.01616, https://doi.org/10.48550/arxiv.1606.09282, https://doi.org/10.48550/arxiv.1904.08039},
to prevent catastrophic forgetting, this approach adds an auxiliary loss to minimize the KL Divergence between the model being trained ($\ad[M]$) and the original initialization ($\ad[M]^*$) as in~\cref{eq:kldiv}:
\begin{equation}
\label{eq:kldiv}
    \ad[L](\ad[M], x,y) = \ctc(\ad[M],x,y) + \lambda \kldiv(p(x), p^*(x))
\end{equation}
where $p(x) = \softmax(\ad[M](x)/T)$ and $p^*(x) = \softmax(\ad[M]^*$ $(x)/T)$ i.e. temperature-scaled logits. We set $\lambda=8.0$ and ablate this choice in Section~\ref{kdablation}. We set $T=1.0$. The other hyperparameters are the same as Full-FT. This approach is even more computationally expensive than Full-FT, because it involves an extra forward pass.

\noindent \underline{(3) Full-FT-Efficient}: This is an efficient variant of Full-FT that only fine-tunes the top few layers such that the number of parameters being fine-tuned is approximately equal to that in DisentangledCL. We finetune $2$ layers for FF and $1$ layer for Att and Conv.

\noindent \underline{(4) KD-Efficient}: This is an efficient variant of the Knowledge Distillation approach, similar to Full-FT-Efficient.

\section{Results}
\label{sec:results}
All results are reported for Librispeech test-clean, test-other and \benchmark{} test sets using both Viterbi and n-gram LM decoding.
\subsection{Evaluating general ASR-trained models}
We first investigate the setting where there is no continual learning performed (i.e. no speaker data is available). Thus, we directly compare a NetAug-trained DisConformer with a baseline model, both trained on LibriSpeech. We report results for LibriSpeech and LibriContinual in Table~\ref{tab:0hr} where all WERs are median WERs across speakers. We run inference on just the DisConformer core ($W_{core}$) discarding all augment experts. This is fair since the DisConformer core and the baseline have the exact same architecture.
We observe that all $3$ DisConformer models consistently outperform the baselines. With LM decoding, DisConformers achieve an average relative WER reduction of $5.6\%$ on LibriSpeech test-clean, $3.7\%$ on test-other and $5.5\%$ on LibriContinual test. This shows that NetAug ASR training is well-suited for obtaining better general-purpose ASR models, even outside the context of continual learning.
\begin{table}[th]\centering
\setlength{\tabcolsep}{4pt}
\scriptsize
\begin{tabular}{rrrrrrrrr}\toprule
& \multicolumn{4}{c}{Viterbi} &\multicolumn{4}{c}{n-gram LM} \\
\cmidrule(lr){2-5}\cmidrule(lr){6-9}
&\multicolumn{2}{c}{LibriSpeech} &\multicolumn{2}{c}{LibriContinual} &\multicolumn{2}{c}{LibriSpeech} &\multicolumn{2}{c}{LibriContinual} \\
\cmidrule(lr){2-3}\cmidrule(lr){4-5}\cmidrule(lr){6-7}\cmidrule(lr){8-9}
Model &test-c &test-o &val &test &test-c &test-o &val &test \\\midrule
Base-FF &5.71 &14.35 &11.46 &12.14 &4.02 &10.16 &7.92 &8.36 \\
DisCo-FF &\textbf{5.38} &\textbf{13.69} &\textbf{10.8} &\textbf{11.22} &\textbf{3.75} &\textbf{9.82} &\textbf{7.41} &\textbf{7.82} \\
\midrule
Base-Att &4.33 &11.23 &8.94 &9.52 &3.42 &8.54 &6.40 &6.76 \\
DisCo-Att & \textbf{4.02} & \textbf{10.76} & \textbf{8.31} & \textbf{8.74} & \textbf{3.29} & \textbf{8.22} & \textbf{6.08} & \textbf{6.34} \\
\midrule
Base-Conv &4.28 &11.31 &9.48 &9.80 &3.50 &8.62 &6.88 &7.22 \\
DisCo-Conv & \textbf{4.13} & \textbf{10.83} & \textbf{8.93} & \textbf{9.36} & \textbf{3.28} & \textbf{8.19} & \textbf{6.66} & \textbf{6.94} \\
\bottomrule
\end{tabular}
\caption{Results on LibriSpeech and \benchmark{} for general ASR-training without Continual Learning.
}\label{tab:0hr}
\end{table}

\subsection{Evaluating Continual Learning}
\label{subsec:mainresults}

\begin{table*}[ht]\centering
\scriptsize
\begin{tabular}{rrrrrrrrrrrrrrrr}\toprule
& & &\multicolumn{6}{c}{Finetuned on 1hr} &\multicolumn{6}{c}{Finetuned on 10hr} \\\cmidrule(lr){4-9}\cmidrule(lr){10-15}
& & &\multicolumn{3}{c}{Viterbi} &\multicolumn{3}{c}{n-gram LM} &\multicolumn{3}{c}{Viterbi} &\multicolumn{3}{c}{n-gram LM} \\\cmidrule(lr){4-6}\cmidrule(lr){7-9}\cmidrule(lr){10-12}\cmidrule(lr){13-15}
& & &\multicolumn{2}{c}{LS} &LC &\multicolumn{2}{c}{LS} &LC &\multicolumn{2}{c}{LS} &LC &\multicolumn{2}{c}{LS} &LC \\\cmidrule(lr){4-5}\cmidrule(lr){7-8}\cmidrule(lr){10-11}\cmidrule(lr){13-14}
Model &CL Algo &\# CL-$\theta$ &test-c &test-o &test &test-c &test-o &test &test-c &test-o &test &test-c &test-o &test \\\midrule
\multirow{4}{*}{Base-FF (16.1M)} &Full-FT &16.1M &8.2 &19.3 &\textbf{9.7} &5.4 &14.0 &\textbf{6.3} &10.8 &26.6 &8.3 &7.2 &20.1 &\textbf{5.6} \\
&KD &16.1M &7.7 &18.3 &9.8 &5.3 &13.5 &6.7 &9.5 &23.4 &\textbf{8.1} &5.0 &17.7 &5.8 \\
&Full-FT-Eff &2.0M &7.5 &17.1 &12.4 &4.8 &11.8 &8.2 &7.8 &17.6 &11.3 &5.0 &12.3 &7.5 \\
&KD-Eff &2.0M &7.2 &16.6 &12.2 &4.8 &11.8 &8.3 &7.6 &17.2 &11.2 &5.0 &12.3 &7.7 \\
\cdashlinelr{1-15}
DisCo-FF (18.3M) &DisCL &2.1M &\textbf{\underline{5.4}} &\textbf{\underline{13.7}} &\underline{10.5} &\textbf{\underline{3.8}} &\textbf{\underline{9.8}} &\underline{6.9} &\textbf{\underline{5.4}} &\textbf{\underline{13.7}} &\underline{9.0} &\textbf{\underline{3.8}} &\textbf{\underline{9.8}} &\underline{6.2} \\
\midrule
\multirow{4}{*}{Base-Att (26.6M)} &Full-FT &26.6M &6.4 &16.0 &7.9 &4.6 &12.0 &\textbf{5.4} &8.6 &22.9 &\textbf{6.9} &6.1 &17.8 &\textbf{4.8} \\
&KD &26.6M &5.9 &14.9 &7.8 &4.3 &11.5 &5.6 &7.3 &19.3 &6.7 &5.3 &14.9 &4.9 \\
&Full-FT-Eff &1.8M &5.3 &12.9 &10.2 &3.8 &9.4 &6.8 &5.4 &13.2 &9.4 &3.9 &9.6 &6.4 \\
&KD-Eff &1.8M &5.2 &12.6 &9.7 &3.8 &9.4 &6.8 &5.3 &12.9 &9.2 &3.8 &9.6 &6.5 \\
\cdashlinelr{1-15}
DisCo-Att (28.7M) &DisCL &2.1M &\textbf{\underline{4.0}} &\textbf{\underline{10.8}} &\textbf{\underline{7.6}} &\textbf{\underline{3.3}} &\textbf{\underline{8.2}} &\underline{5.5} &\textbf{\underline{4.0}} &\textbf{\underline{10.8}} &\underline{7.0} &\textbf{\underline{3.3}} &\textbf{\underline{8.2}} &\underline{4.9} \\
\midrule
\multirow{4}{*}{Base-Conv (27M)} &Full-FT &27M &6.5 &16.4 &8.0 &4.7 &12.3 &\textbf{5.5} &8.8 &23.4 &7.0 &6.2 &18.1 &\textbf{4.9} \\
&KD &27M &6.0 &15.3 &7.9 &4.5 &11.9 &5.7 &7.4 &19.7 &\textbf{6.9} &5.4 &15.4 &5.1 \\
&Full-FT-Eff &1.7M &5.4 &13.1 &10.7 &3.9 &9.5 &7.1 &5.5 &13.5 &9.9 &3.9 &9.8 &6.7 \\
&KD-Eff &1.7M &5.2 &12.9 &10.3 &3.9 &9.6 &7.2 &5.3 &13.2 &9.6 &3.9 &9.8 &6.8 \\
\cdashlinelr{1-15}
DisCo-Conv (28.3M) &DisCL &1.3M &\textbf{\underline{4.1}} &\textbf{\underline{10.8}} &\textbf{\underline{7.8}} &\textbf{\underline{3.3}} &\textbf{\underline{8.2}} &\textbf{\underline{5.5}} &\textbf{\underline{4.1}} &\textbf{\underline{10.8}} &\textbf{\underline{6.9}} &\textbf{\underline{3.3}} &\textbf{\underline{8.2}} &\textbf{\underline{4.9}} \\
\bottomrule
\end{tabular}
\caption{Results on LibriSpeech and \benchmark{} with Viterbi and n-gram LM deocding. All WERs are median WERs across speakers. ($x$M) next to model in parentheses denotes total model params. \# CL-$\theta$ is the \# available params for CL. LS=LibriSpeech, LC=LibriContinual. \textbf{Bold} numbers are the best WERs across all approaches. \underline{Underlined} numbers are the best WERs across \# CL-$\theta$-matched approaches. }\label{tab:mainresults}
\end{table*}
Table~\ref{tab:mainresults} presents the evaluation results on the \benchmark{} benchmark.  
As the amount of speaker data increases ($0$-hr from Table~\ref{tab:0hr} to $1$ hr to $10$ hr), the WERs on the \benchmark{} val/test sets decrease as expected.

\noindent \textbf{Performance on LibriSpeech.}
We start by analyzing preservation of performance on the original dataset, Librispeech, after continual learning. The performance of all baselines degrades considerably, with the effect more pronounced for the $10$ hr split, exhibiting catastrophic forgetting. KD performs better than Full-FT (likely due to the KL-divergence term), and the Efficient variants perform better than the vanilla approaches (likely since only a subset of parameters are finetuned). In contrast, our DisentangledCL has the same performance as the general ASR model from Table~\ref{tab:0hr} , resulting in no catastrophic forgetting at all and it significantly outperforms the best baseline. Averaged across all settings, over the best baseline, this results in relative WER gains of $23.94\%$ with Viterbi and $17.66\%$ with n-gram LM for test-clean, and $17.06\%$ and $15.58\%$ respectively for test-other.

\noindent \textbf{Performance on LibriContinual.}
On LibriContinual, in all settings, the DisConformer models with at most $13\%$ extra total params significantly outperform the \#CL-Param-matched Efficient baselines.
Averaged across all settings, over the best Efficient baseline, this is a relative WER gain of $21.16\%$ with Viterbi and $18.26\%$ with n-gram LM for the LibriContinual validation set, and $21.85\%$ and $20.65$ respectively for the test set.
Surprisingly, in both the $1$ hr and $10$ hr settings, despite only finetuning a much smaller fraction of parameters ($7\%$ at maximum) the DisConformer-Att and DisConformer-Conv are within $\pm 0.1$ WER of fully finetuned baselines (Full-FT and KD). In contrast, the DisConformer-FF model performs much worse than the fully finetuned baselines, likely owing to the much smaller number of trainable params. It has a max abs.\@ WER difference of $+0.9$ with the best baseline Full-FT across all settings. On the other hand, on LibriSpeech, Full-FT has a much worse WER performance; min $-4.7$ abs.\@ WER across all settings. This is a tradeoff between speaker-specific and general performance. Depending on the end use-case, the magnitude of acceptable degradation of general vs. speaker-specific performance will vary. 

Overall, this analysis reveals that with a small number of available parameters for finetuning (at most $13\%$ of baselines), the DisConformer models offer superior performance on Librispeech and on speaker-specific LibriContinual, they perform better than trainable-parameter-matched baselines, and are sometimes able to match even fully-finetuned baselines (for Att and Conv, but not FF). This also suggests that DisConformers may be more effective when applied to Att or Conv layers than FF.

\subsection{Ablations}
\label{sec:ablation}
\label{subsec:ablation1}
\noindent \textbf{Using DisentangledCL on baseline models.} We analyze whether NetAug disentangled training is necessary by training baseline models in the disentangled `core+augment' framework. We perform LM-decoded LibriContinual test set eval with the $1$ hr train set for these $4$ settings which all have the same architectures:

\noindent \underline{Base-Conv + Random.} We use the Base-Conv model as the core and randomly initialize $k_{conv}=12$ augment experts for finetuning.

\noindent \underline{Base-Conv + Base-Conv.} We train a Base-Conv Conformer with $224$ conv channels with CTC loss and treat its first $128 = 8 {\times} 16$ channels as the core and the last $96 = 8 {\times} 12$ channels as the augment experts.

\noindent \underline{DisCo-Conv (core) + Random.}  We take the trained DisConformer -Conv and randomly re-initialize its augment parameters.

\noindent \underline{DisCo-Conv (core) + Disco-Conv.} This is our DisCo-Conv model.
We find that these $4$ settings achieve a test set WER of (a) $5.88$, (b) $5.72$, (c) $5.71$ and (d) $5.52$ respectively. NetAug-trained core experts are better [(a) $5.88$ vs.\@ (c) $5.71$] and NetAug-trained augment experts are better [(c) $5.71$ vs.\@ (d) $5.52$], showing NetAug is important. (b) $5.72$ shows that baselines can also be trained in our DisentangledCL framework, although not as well as our model, (d) $5.52$.

\begin{figure}[t]
    \centering
    \includegraphics[width=\columnwidth]{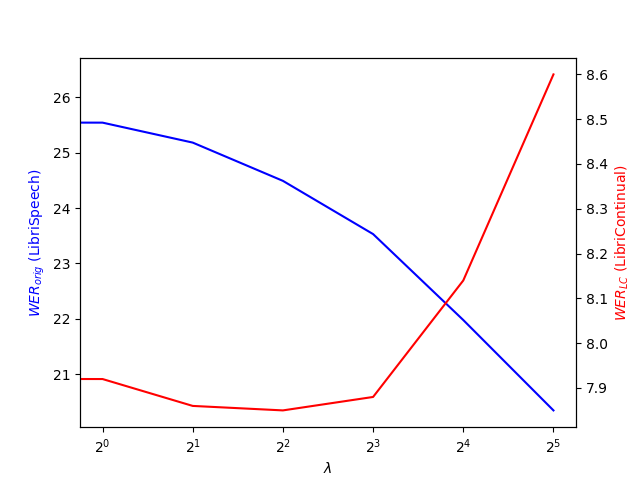}
    \caption{Ablating the KL-divergence weight parameter $\lambda$ for the KD baselines.}
    \label{fig:kdablation}
\end{figure}

\label{kdablation}
\noindent \textbf{Ablating KD hyperparam $\lambda$.} The $\lambda$ hyperparam in the KD baseline loss controls the tradeoff between learning on new data (CTC loss) and staying close to the old model (KL div loss). We tune $\lambda$ over the set $\{0,1,2,4,8,16,32\}$ using the Base-FF model finetuned on the $10$ hr split and decoded with Viterbi. 
Figure~\ref{fig:kdablation} depicts the LibriSpeech and LibriContinual performance for different values of $\lambda$.
As $\lambda$ increases, performance on LibriSpeech monotonically decreases (from $26.4$ at $\lambda {=} 0$ to $20.2$ at $\lambda {=} 32$); however, while LibriContinual performance is also improved from $8.2$ at $\lambda {=} 0$ to $7.9$ at $\lambda {=} 8$, it significantly worsens (to $8.6$ at $\lambda {=} 32$). Thus, we choose $\lambda=8$ for the KD baseline.

\section{Conclusion}
We introduced \benchmark{}, a new continual learning benchmark for efficient speaker-specific domain adaptation. We also proposed DisConformers and novel ASR training (NetAug) and continual learning (DisentangledCL) algorithms which use different parts of the same model to achieve strong general ASR performance and speaker-specific performance in a parameter-efficient manner. For future work, we plan to extend the LibriContinual benchmark to the unlabelled setting (via weak supervision for ASR) and add more speech tasks. We also plan to extend our NetAug algorithm to build speaker-specialized experts.

\section{Acknowledgements}
The research platform for this work was built on top of \cite{torchaudio}. In addition to the support on guidance of torchaudio components, we are thankful for the contribution from Xiaohui Zhang and Zhaoheng Ni from Meta AI for their technical suggestions and collaborations.

\bibliographystyle{IEEEbib}
\bibliography{strings,refs}

\end{document}